\documentclass{aa}
\usepackage[dvips]{graphicx}

\usepackage{txfonts}
\usepackage{natbib}
\usepackage[latin1]{inputenc}

\usepackage[dvips]{graphicx}
\usepackage{sp}

\renewcommand{\na}{\nabla}

\begin{document}
\title{On the power spectrum of solar surface flows}

\author{M. Rieutord\inst{1}, T. Roudier\inst{1}, F. Rincon\inst{1},
J.-M. Malherbe\inst{2}, N. Meunier\inst{3}, T. Berger\inst{4} and Z. Frank\inst{4}}

\authorrunning{Rieutord et al.}

\institute{
Laboratoire d'Astrophysique de Toulouse-Tarbes,
Universit\'e de Toulouse, CNRS, 14 avenue E. Belin, 31400 Toulouse, France
\and LESIA, Observatoire de Paris, Section de Meudon, 92195 Meudon, France
\and LAOG, Universit\'e Joseph Fourier, CNRS, BP 43, 38041 Grenoble Cedex, France
\and Lockheed Martin Advance Technology Center, Palo Alto, CA, USA
\\
\email{\tiny [rieutord,roudier,rincon]@ast.obs-mip.fr, Jean-Marie.Malherbe@obspm.fr,
nmeunier@obs.ujf-grenoble.fr,\\ berger@lmsal.com, zoe@lmsal.com}
}
\date{\today}

\offprints{M. Rieutord}

\abstract{% Context
The surface of the Sun provides us with a unique and very detailed view 
of turbulent stellar convection. Studying its dynamics can therefore 
help us make significant progress in stellar convection modelling. 
Many features of solar surface turbulence like the supergranulation
are still poorly understood.
}{%Aims
The aim of this work is to give new observational constraints on these 
flows by determining the horizontal scale dependence of
the velocity and intensity fields, as represented by their power spectra,
and to offer some theoretical guidelines to interpret these spectra.
}{%Method
We use long time-series of images taken by the Solar Optical
Telescope (SOT) on board the Hinode satellite; we reconstruct both 
horizontal (by granule tracking) and vertical (by Doppler effect)
velocity fields in a field-of-view of $\sim 75\times75$~Mm$^2$. The
dynamics in the subgranulation range can be investigated with
unprecedented precision thanks to the absence of seeing effects and the
use of the modulation transfer function of SOT for correcting the spectra.
}{% Results
At small subgranulation scales down to 0.4~Mm the 
spectral density of kinetic energy associated with vertical motions 
exhibits a $k^{-10/3}$-like power law, while the intensity fluctuation spectrum
follows either a $k^{-17/3}$ or a $k^{-3}$-like power law at the
two continuum levels investigated (525 and 450~nm respectively). 
We discuss the possible physical origin of these scalings 
and interpret the combined presence of $k^{-17/3}$ and  $k^{-10/3}$ 
power laws for the intensity and vertical velocity as a signature of 
buoyancy-driven turbulent dynamics in a strongly thermally diffusive regime.
In the mesogranulation range and up to a scale of 25~Mm, we find that the
amplitude of the vertical velocity field decreases like $\lambda^{-3/2}$ with
the horizontal scale $\lambda$. This behaviour corresponds to a $k^2$
spectral power law. Still in the 2.5--10~Mm mesoscale range, we find
that intensity fluctuations in the blue continuum also follow a $k^2$ 
power law. In passing we show that granule tracking cannot sample scales
below 2.5~Mm.  We finally further confirm the presence of a significant
supergranulation energy peak at 30~Mm in the horizontal velocity power
spectrum and show that the emergence of a pore erases this spectral peak.
We tentatively estimate the scale height of the vertical velocity field
in the supergranulation range and find 1~Mm; this value suggests that
supergranulation flows are shallow.

}{% Conclusions
}

\maketitle

\section{Introduction}

A complete understanding of thermal convection in stars remains one of the
main challenges of present day astrophysics. The Sun offers an unsurpassed
detailed view to address this question.  Solar convection is
highly non-linear -- typical Reynolds numbers are over $10^{10}$ -- making
fully detailed direct numerical simulations down to the viscous dissipation 
scales unaffordable. In spite of this difficulty, some observed features of 
solar convection like for instance granulation
are rather well understood \citep[e.g.][]{SN00}. Dynamical features at larger scales 
like the supergranulation pattern are much less understood. 
Supergranulation is characterised by a horizontal velocity field spanning scales from
15~Mm to 75~Mm according to the recent work of \cite{RMRRBP08}.  Despite
several attempts, supergranulation has not been identified in large-eddy
or direct numerical simulations \citep[e.g.][]{RLRNS02,RLR05,SGSNB09}. Its
origin remains unknown, although some scenarios have been proposed. The
classical picture \citep{SL64} is a linear thermal convection scenario
which associates supergranulation cells with  the second ionisation of helium.
Other scenarios put forward the idea that supergranulation is a
surface phenomenon, as actually indicated by the recent results of local
helioseismology \citep{GB05}. \cite{RRMR00} suggested that it could be
the result of a large-scale instability of surface granular convection.
\cite{RR03b} proposed that some fixed flux boundary condition is
imposed by the granulation to the layers just below: with these boundary conditions, 
the buoyancy destabilises the largest available scales. The associated convective 
motions have a fairly low intensity contrast \citep{vdb74}, very much like 
supergranulation cells \citep{MRT07}. \cite{RR03b} further showed that an 
effective finite but large horizontal convection scale can be obtained in 
this framework by considering the dynamical 
effects of a mean magnetic field on the flow. 
Collective plume interactions \citep{R03a,CCT07} and travelling-wave 
sheared \citep{GK06} convection or magnetoconvection \citep{GK07} 
have also been suggested as a possible origin of supergranulation.

Overall the constraints on the large-scale dynamics of solar
convection imposed by models and theories are still very
loose. Guidance from observations is thus much desired. Particularly
if a dynamical connection exists between
granulation and supergranulation, some hints of such a 
scale interaction should be found in the dynamics of intermediate
horizontal scales. Flows in the 3--10~Mm range
are little known. This range of scales is traditionally referred to as
the mesogranulation range after the work of \cite{NTGS81}, but the very
existence of genuine enhanced convective motions at these scales has been 
much debated \citep{SDF92,SB97,RRMR00,SSH00}. As pointed out in \cite{RRMR00},
the main problem with mesogranulation was its way of detection, namely
through the measurement of the horizontal divergence of the flows, which
is highly sensitive to the way data are reduced. The very nature of the
flows at these scales is nonetheless important to answer the question
of the origin of supergranulation.

In order to make progress, it is clear that the large-scale side of the
granulation spectral peak should be investigated in more detail. Simple
models like the mixing-length theory or plume dynamics \citep{RZ97}
do not predict any special spectral feature when the scale (horizontal
or vertical) grows. Hydrodynamic numerical simulations like the recent
ones of \cite{SGSNB09} seem to go in the same direction and do not
exhibit any spectral feature reminiscent of supergranulation \cite[see
also][]{NSA09}.  However, kinetic energy spectra derived from the radial
velocity measured by SOHO/MDI (i.e. from dopplergrams) do show a rise
of the spectral kinetic energy density at a scale of 11~Mm and a peak
at 36~Mm \citep{HBBBKPHR00}.  These results are clearly confirmed
by the recent measurement of the horizontal surface flows at disc
centre with the wide-field high resolution camera CALAS at Pic-du-Midi
\citep{RMRRBP08}. These observations are therefore clearly at odds
with the most advanced numerical models of solar surface convection.

\begin{figure}[t]
\centerline{\includegraphics[width=\linewidth]{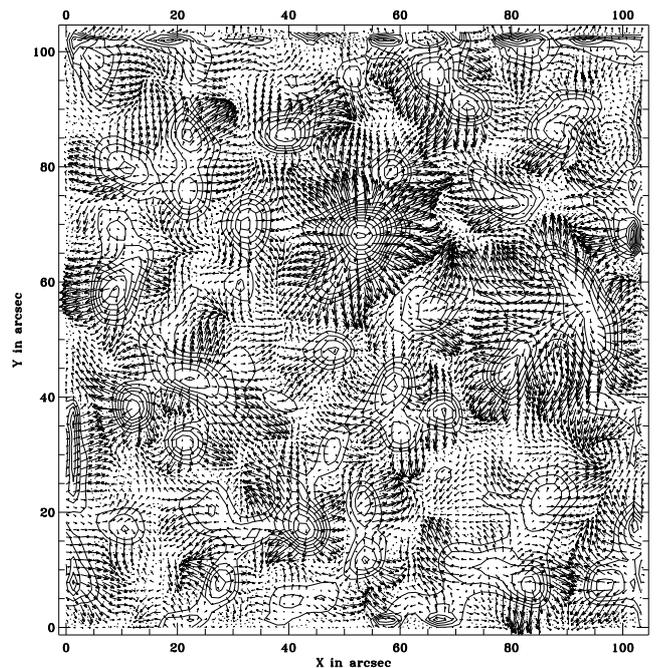}}
\caption[]{Velocity field derived from the motion of granules during a
45~min window, interpolated by a Daubechies wavelet of
4*953~km. Contours mark the horizontal divergence of the velocity.}
\label{velo}
\end{figure}

\begin{figure}[t]
\centerline{\includegraphics[width=\linewidth]{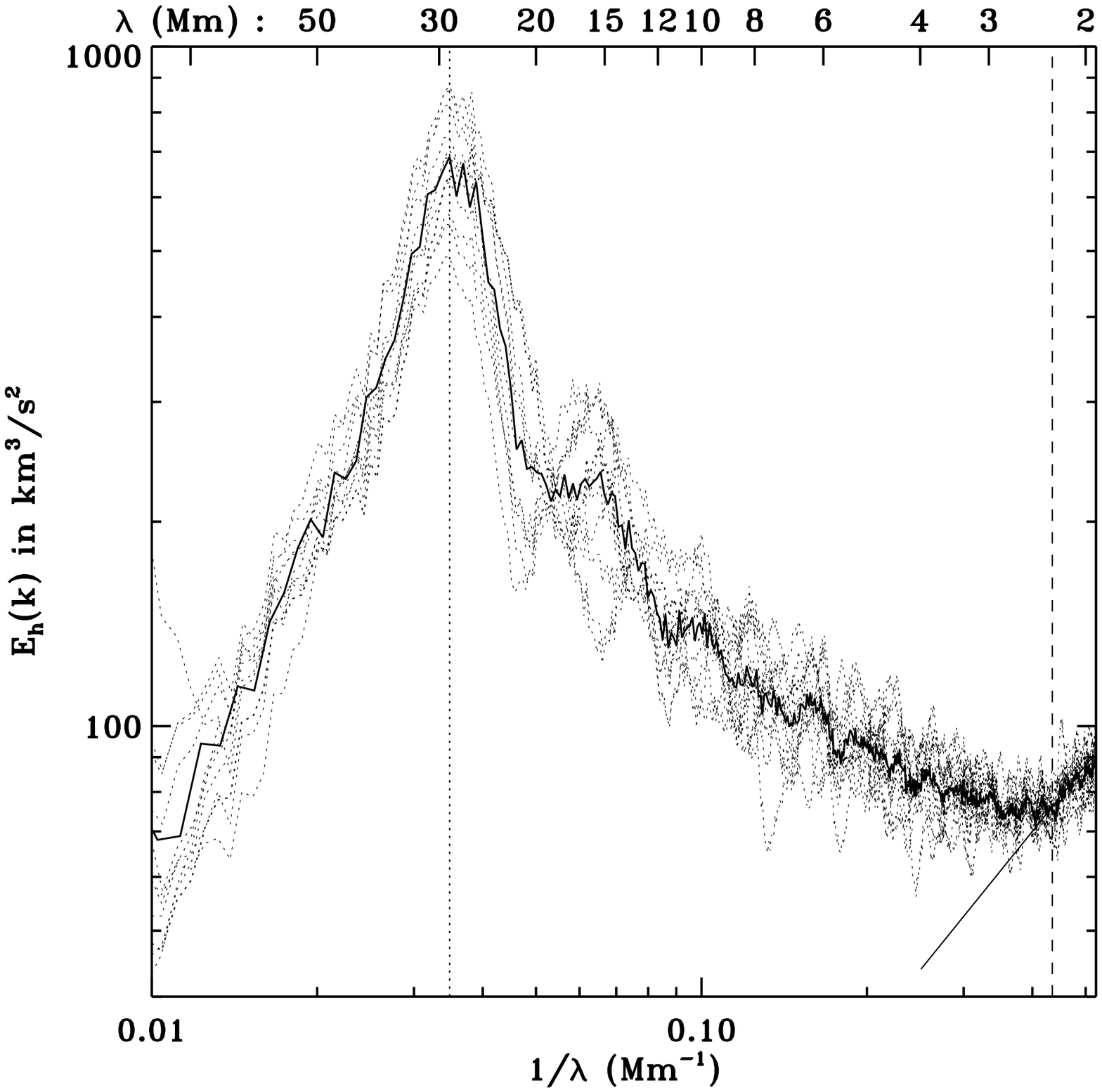}}
\caption[]{Mean spectra and fluctuations: eleven independent
horizontal kinetic energy spectra (dotted lines) computed with a
three-hours time window, along with their average (thick solid line)
are represented.
Note that the fluctuations may reach an amplitude of 40\% compared to
the mean value. The straight line on the right indicates the power law
$E(k)\sim k$ characteristic of decorrelated noise. The vertical 
straight line marks the 2.3~Mm scale. Here $N_x=N_y=80$ and $N_p=1024$.
}
\label{mean_spec}
\end{figure}

In this paper we wish to provide accurate observational facts
to guide modelling efforts and impose more constraints for theories.
For this purpose we use new data sets collected
by the Hinode mission and especially by the SOT instrument. Because
of their high spatial resolution and their very low noise, these data
are appropriate to either refine previous measurements or investigate
new properties of the flows. In particular the subgranulation range can now
be investigated as never before. We shall focus our attention on the
power spectra of the flows because the information that
can be obtained from such representations is usually the
most robust aspect of these highly turbulent flows. 

The paper is organised as follows: after a brief presentation of the data 
and the techniques to process them (Sect.~\ref{sectiontwo}), 
we detail the spectral properties of the
measured velocity fields (Sect.~\ref{sectionthree}) and
the intensity field (Sect.~\ref{sectionfour}).  A discussion and physical
interpretation of the results follows (Sect.~\ref{discussion}). 
The main results and conclusions are summarised at the end of the paper
(Sect.~\ref{conc}).

\begin{table*}[t]
\caption[]{The position and amplitude of the supergranulation peak for various data sets.}
\centering \vspace{1mm}
\begin{tabular}{ccccccc} \hline
Date & Instrument & FOV (Mm$^2$)& Filter & Time slot & $\lambda_{\rm
SG}$ (Mm) & $E_{\rm SG}$ (km$^3/s^2$) \\
\multicolumn{7}{c}{ Wide-field data} \\
1996 May-Jul & SOHO/MDI & global & Ni I, 676.8~nm & 62d & 36.4 & (200)\\
2000 Apr 22& TRACE & 225x250 & WL & 6x3 h &  35.6 & 200 \\
2007 Apr 24& TRACE & 238x275 & WL & 2x3.4 h &  29.8 & 490 \\
2007 Mar 13& CALAS & 290x216 & 575~nm & 2x3 h & 36.4 & 350 \\
\multicolumn{7}{c}{Small-field data} \\
2007 Mar 8& Hinode/SOT& 75.9x76.8  & G-band &  4h   & no peak   & 220 \\
2007 Mar 15& Hinode/SOT& 77.3x33.9 & G-band &  3x3h & 57/17-27  & 240/350  \\
2007 Aug 29-30& Hinode/SOT & 76.6x76.6 & blue & 11x3h & 28.7 & 600 \\
2007 Oct 1 & Hinode/SOT & 2x(76x76) & blue & 2x3.8h & 27/31 & 460/545 \\
\hline
\end{tabular}
\label{SG_table}
\end{table*}

\begin{figure*}[ht]
\centerline{\includegraphics[width=0.5\linewidth]{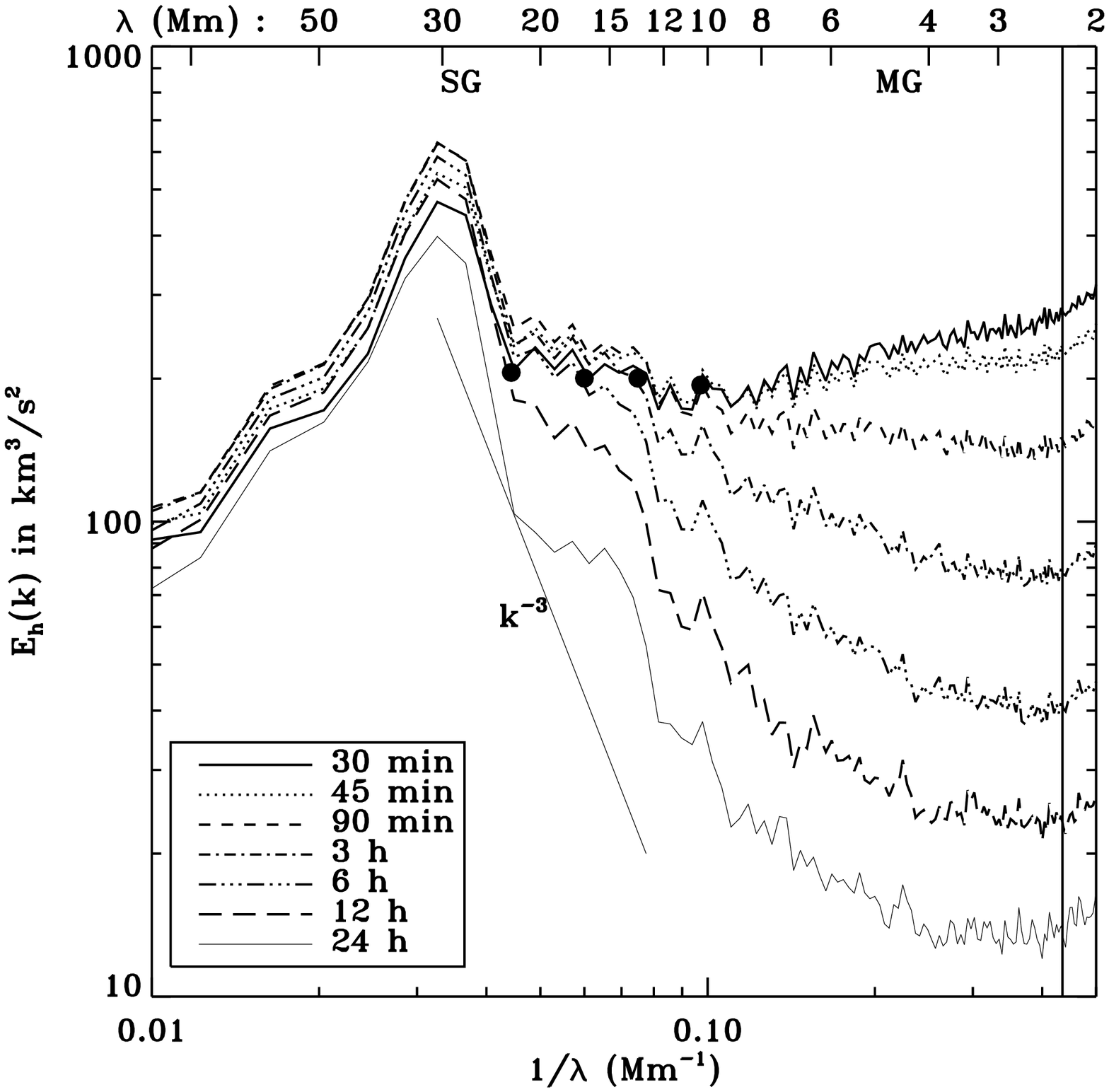}
\includegraphics[width=0.5\linewidth]{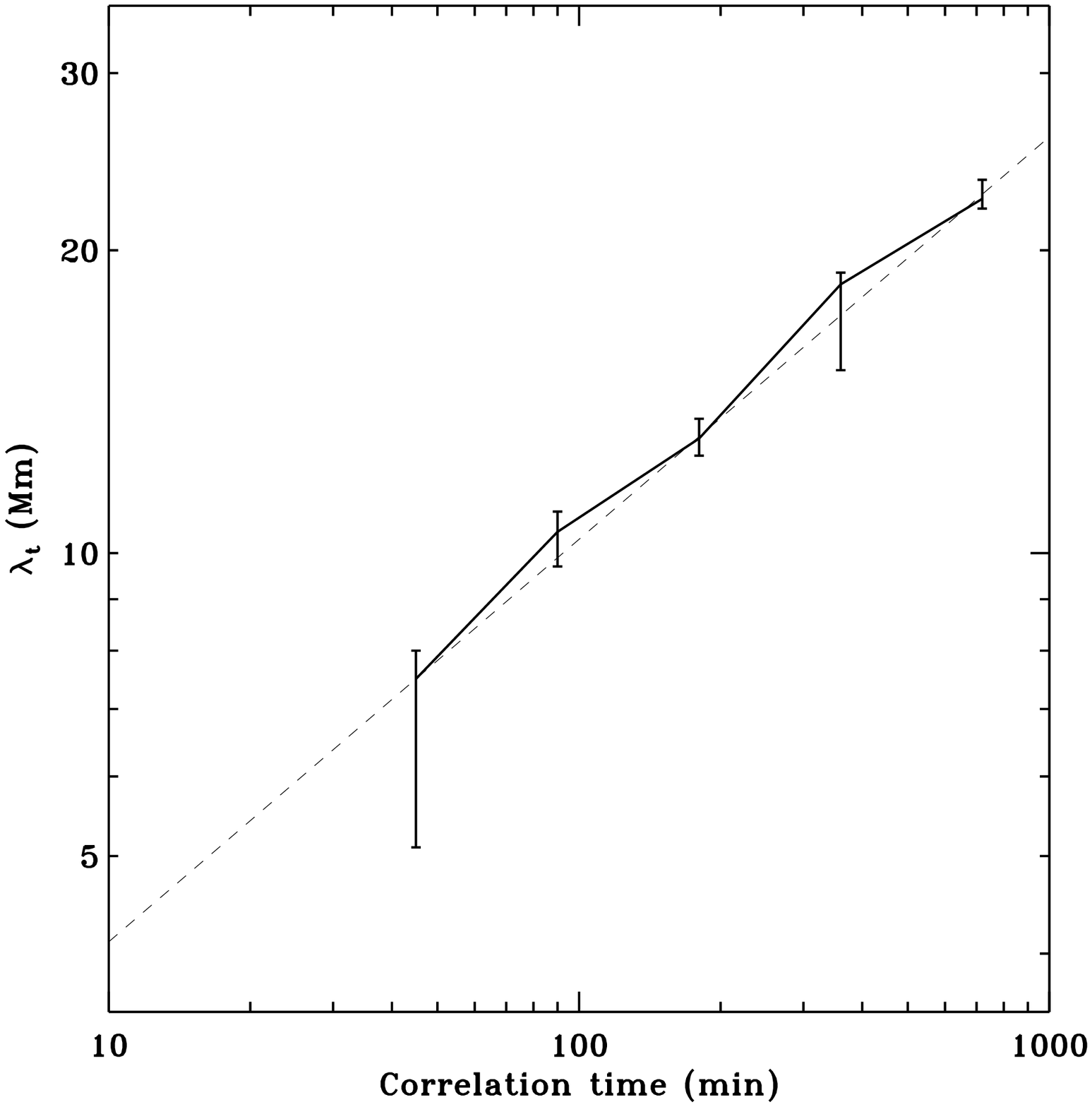}}
\caption[]{Left: kinetic energy spectra of the horizontal velocity field
for various time windows. The vertical line marks the 2.3~Mm scale below
which noise dominates. The black dots mark the scale at which the spectrum
of a given time average disconnects from the 30-min one. Right: relationship
between the length scale and the correlation time scale for the horizontal
velocity field. Here $N_x=N_y=80$ and $N_p=256$.}
\label{KES}
\end{figure*}

\section{The data and the reduction procedure\label{sectiontwo}}

We used multi-wavelength data sets of the Solar Optical Telescope (SOT)
on board the HINODE\footnote{The Hinode spacecraft launched in 2006
was designed and is now operated by JAXA (Japanese Space Authorities)
in cooperation with NASA (National Aeronautics and Space Administration)
and ESA (European Space Agency).} mission \citep[e.g.][]{ITSSO04,STISO08}.
The SOT has a 50~cm primary mirror with a spatial resolution about
0.2\arcsec at 550~nm.

\subsection{Horizontal velocities}\label{vh}

A first set of images taken at disc centre was recorded continuously
from 29 August 10:17 UT until 31 August 10:19 UT 2007, except for an
interruption of seven minutes on 30 August at 10:43 UT. We used the set taken with the
broadband filter imager (BFI) with a spectral width of 0.22~nm 
\citep{WR09} in the blue continuum at 450.45~nm.  The mean time step 
between two successive frames is 50.1 sec. The field-of-view with BFI 
is $111\farcs6\times111\farcs6$
with a pixel of 0\farcs109 ($1024\times1024$). After alignment,
the useful field-of-view was reduced to $104\farcs6\times104\farcs6$ or
$76.3\times76.3$~Mm$^2$ during 33h.  The data were also $k-\omega$
filtered with a phase velocity threshold of 6 km~s$^{-1}$ to keep only
convective motions.

We determined the horizontal velocity field for various time averages
from this set of data, using our algorithm of granule tracking CST
\citep{RRRD07}. We recall that because of the nature of granules their
motion is representative of the large-scale plasma flow only for length
scales larger than 2.5~Mm and time scales longer than 30~min. These
limits have been derived from numerical simulations by \cite{RRLNS01}
and are confirmed by these data (see below).

Other data sets were processed in the same way to compare
several flow characteristics (see Table 1).

\subsection{Vertical velocities}\label{vv}

To determine the vertical component of the velocity we used a second data 
set from SOT/Hinode, also taken at disc centre
and recorded on 4 September 2009 21:08 UT to 22:25 UT. These data were
taken with the narrowband filter imager (NFI) with a spectral width of
9~pm, near the FeI line at $\lambda_0=$557.6~nm, sampling nine wavelengths
in the line profile. This regular sampling allowed us to accurately measure
the Doppler shift of the line every 28.5s. The field-of-view 
of this series is $82\arcsec\times82\arcsec$, which 
represents $60\times60$~Mm$^2$. The pixel size is $0\farcs08$.

\begin{figure}[t]
\centerline{\includegraphics[width=0.9\linewidth]{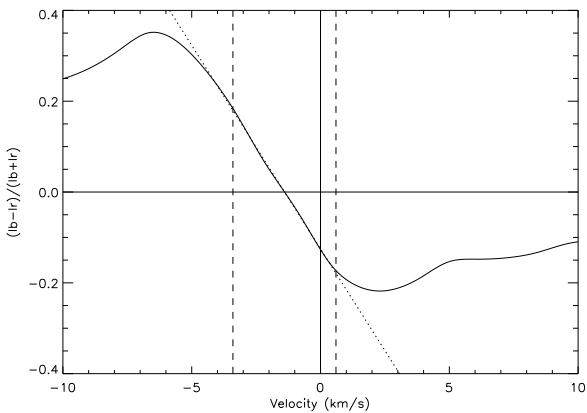}}
\caption[]{Relationship between radial velocities and relative intensity
fluctuations, using the two wavelengths at -1.8~pm and +10.8~pm aside the FeI
line at 525~nm. Positive velocities are towards the observer.}
\label{calib}
\end{figure}

These data are completed by a third data set from SOT/Hinode, taken at
disc centre and recorded on 4 September 2009 7:35 UT to 10:17 UT with
a regular cadence of one image every 57.5~s. These data were also obtained 
with the NFI but around the FeI line at 525.0208~nm. However, in that case
the line profile was sampled only at two wavelengths, 
namely $\lambda_0-1.8$~pm and $\lambda_0+10.8$~pm, which did not 
allow us to determine the Doppler shift of the line in the same way 
as with the second data set. The radial velocity was instead estimated 
through the simple relationship

\beq v=k\frac{I_b-I_r}{I_b+I_r} + v_0\, ,\eeqn{calib_eq}
where $v_0$ and $k$ are constants and $I_r$ and $I_b$ are the intensities
in red and blue positions of the filter ($+10.8$~pm and $-1.8$~pm)
respectively. The actual relationship between the radial velocity and the
intensity difference for this particular line and positions of the filter
is shown in Fig.~\ref{calib}. Note that an unavoidable pitfall with NFI is
that the red and blue positions of the filter are in general asymmetric
with respect to $\lambda_0$. From the previous plot we see that this
implies that large velocities cannot be measured correctly on one side
of the line for these data (here on the blue side). The field-of-view of
this series is $112\farcs6\times112\farcs6$ with a pixel of $0\farcs16$,
corresponding to a physical field of $82\times82$~Mm$^2$.

The Doppler fields were then $k-\omega$ filtered to remove fast horizontal
motions of the patterns with a threshold of 6 km~s$^{-1}$.

Unfortunately no flat-field image was available for the data sets
taken at $\lambda_0=$557.6~nm. We tried to circumvent this problem
by creating a flat-field image for each image. For this purpose
we averaged 50 images around the desired one and smoothed the result
with a window of 16\arcsec$\times$16\arcsec. The operation possibly
altered the signal at scales larger than 12~Mm. Note finally that unlike
the $\lambda_0=$557.6~nm data sets,  we were able to correct the data
set taken around $\lambda_0=$525.0~nm with a genuine flat-field image.

Finally, following \cite{Danilovic_etal08} we computed the PSF and MTF
of SOT to deconvolve all the images including those used to derive the
Doppler shift from the instrument effects (central obscuration, spider
and ccd).  This correction turned out to be crucial for the investigation
of subgranulation-scale turbulence, which is one of the novelties brought
by the Hinode mission.

\section{Velocity fields\label{sectionthree}}

\subsection{Horizontal velocity fields}

We give in Fig.~\ref{velo} a typical view of the velocity field
using a time average of 45~min, on which the supergranulation pattern
is clearly visible.  As is customary with any turbulent flow, it is
crucial to investigate the spectral signature of these flows.  The high
quality of the data enables us to use the highest spatial and temporal
resolutions available to granule tracking. Using the CST algorithm, as
in \cite{RMRRBP08}, we computed the kinetic energy spectrum $E_h(k)$
for various time windows for the horizontal motions as shown in
Fig.~\ref{mean_spec}. We recall its definition

\[ \demi\moy{\overline{v}_h^2} = \int_0^\infty E_h(k)dk~,\]
where $\overline{v}_h$ denotes the horizontal velocity field averaged
over the time window during which the granule tracking is performed;
$\moy{}$ should be an ensemble average; here it is a spatial average,
assuming that the turbulent flow is ergodic.

For comparisons with other data sets, we plotted the dimensional values
(in km$^3/$s$^2$) of the kinetic energy spectral densities in all the
figures.  Since this value depends on the normalisation of the Fourier
transform, we give the details of the expressions used in the paper in
Appendix~\ref{app}. In order to give a better spectral resolution, all
spectra were computed using zero-padding. Thus, we give the size of the
data ($N_x,N_y$) and the size of the padding area $N_p$ for each spectrum.

\begin{figure}[t]
\centerline{\includegraphics[width=0.9\linewidth]{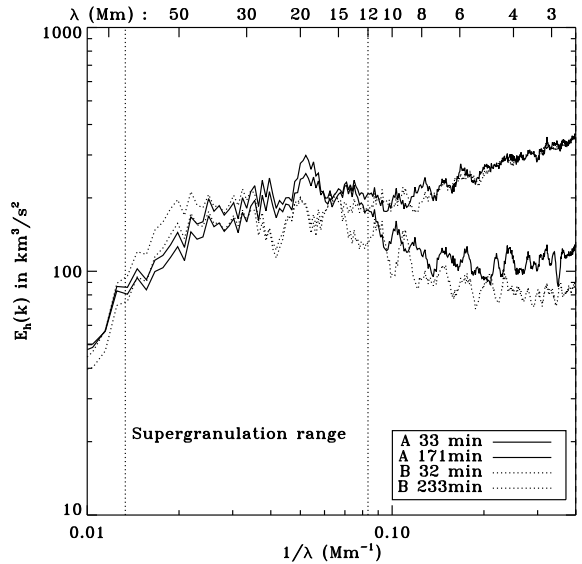}}
\caption[]{Kinetic energy spectra of the horizontal velocity field
during the emergence of two pores in the field-of-view. The solid
line corresponds to the first series, at the end of which one pore
has emerged. The dotted one corresponds to the second series, during
which the second pore emerges.  For each series, two time windows were
used. The vertical dotted lines mark the 12.5--75~Mm range where the
supergranulation peak is found (cf. Fig.~\ref{KES}). Here $N_x=81,
N_y=82$ and $N_p=1024$.
}
\label{pore}
\end{figure}

These spectra are interesting for various reasons. First, we notice
in Fig.~\ref{mean_spec} the prominent supergranulation peak centred
at 30~Mm. This length scale is slightly shorter than the one measured
by \cite{RMRRBP08}, who found the maximum at 36.4~Mm. This change in the
peak wavelength does not seem to be an artefact of the smaller field.
Indeed, reducing the field on the data of \cite{RMRRBP08}
to a size comparable to that of the present data, one still obtains a value of
36.4~Mm. The difference seems to result from an intrinsic variability. To
verify this point, we analysed other data from various sources (other
Hinode fields, some data from the TRACE satellite). The results,
together with those obtained from SOHO/MDI data by 
\cite{HBBBKPHR00}\footnote{The amplitude of the peak for 
these data, given in between parenthesis, is an estimate resulting from a crude 
conversion of Doppler signal to horizontal velocities.}
are summarised in Table~\ref{SG_table}. These data indeed show some
variability both in the location of the peak and its amplitude.
The fact that the small fields-of-view tend to give smaller values of
the supergranulation scale may not be statistically significant.

Figure~\ref{mean_spec} also displays a plot of the individual spectra
together with their average.  These spectra provide further evidence
of the strong variability of the horizontal velocity field even at
the supergranulation scale. They are constructed from three-hours-averaged
velocity fields. We note that they do not show any energy excess within
the mesogranulation range, i.e. between 3 and 10~Mm. The absence of any
specific spectral feature in this range is confirmed when other time
windows are used (Fig.~\ref{KES}).

In Fig.~\ref{KES} (left) we plot the mean spectra obtained with
time windows ranging from 0.5 to 24~hours. Each spectrum is an
average of all the independent spectra that could be extracted from the
whole series. For instance the 0.5h time window allowed us to extract 66
independent spectra from the 33~hrs of data. Note that all spectra show a
distinct rise at scales below 2.3~Mm ($1/\lambda> 0.43$~Mm$^{-1}$). This
rise follows a linear law ($E_h(k)\sim k$) as expected from a
two-dimensional uncorrelated noise. Hence we clearly find the spatial
resolution limit imposed by granule tracking. This confirms the analysis
performed on the numerical simulations by \cite{RRLNS01}, which suggested
that granule tracking could not be used at scales smaller than 2.5~Mm.

The 30-minutes spectrum is the closest to an instantaneous one. This spectrum
displays three features: (i) the supergranulation peak, (ii) a minimum
in the kinetic energy density at 10~Mm and (iii) a slow monotonic increase
towards the small scales.
Feature (ii) was also found by
\cite{HBBBKPHR00} in the Doppler measurements from SOHO/MDI; their
spherical harmonic spectrum shows a minimum near $\ell=400$ which
corresponds to $\lambda=11$~Mm, quite close to our value.

We note that the spectra obtained with time windows ranging from 30~min to
24~h almost superpose at large scales. However,  shifting the attention
to smaller scales, one finds the existence of a scale $\lambda_t$ where
the spectrum disconnects from the 30~min one (marked with black dots in
Fig.~\ref{KES}). The time average acts as a low-pass filter which
removes (or simply weakens) the flows on shorter time scales.  The size
of the time window can hence be used as a proxy of the correlation time
of the flow at the scale marked by the dot. Figure~\ref{KES} (right) shows
the scale $\lambda_t$ as a function of the time average associated with
the time window. We find that the relationship between this length scale
and the window size is close to a power law $\lambda_t\sim
t^{2/5}$. As the correlation time is $\sim \lambda/v_\lambda$,
we can relate $v_\lambda$ and $\lambda$ and find $v_\lambda\sim
\lambda^{-3/2}$. This translates into a power law spectrum $E(k)\sim
k^2$ for the kinetic energy. This exponent is typical of the range
7--25~Mm.

Finally, while processing other data sets to better evaluate the variability of
the supergranulation peak, we came across the data recorded in the G-band
by Hinode/SOT on 8 March 2007. Two series of images taken at [2:55, 5:45]
UT and [6:00, 9:52] UT drew our attention. As shown in Fig.~\ref{pore},
the supergranulation peak has disappeared if we consider the shortest
time sampling. In this figure we distinguished the two series because
one pore is emerging at the end of the first series (around 4:40 UT)
and the second pore emerges in the second series (around 6:18 UT). We
note that the kinetic energy density in the supergranulation range has
weakened at scales shorter than 33~Mm in the second series when the
pair of pores has reached its steady state. We also observe that the TRACE
data taken in April 2000 during the solar maximum also show
a weaker supergranulation peak compared to the analogous data of 2007
taken at the solar minimum. Hence we note a trend that magnetic fields
seem to markedly affect the amplitude of the supergranulation
flows. This is likely connected with the anti-correlation between the
size of the supergranules and the strength of the magnetic field observed
by \cite{MRT07,MRR08}.

\begin{figure*}[htp]
\centerline{
\includegraphics[width=0.48\linewidth]{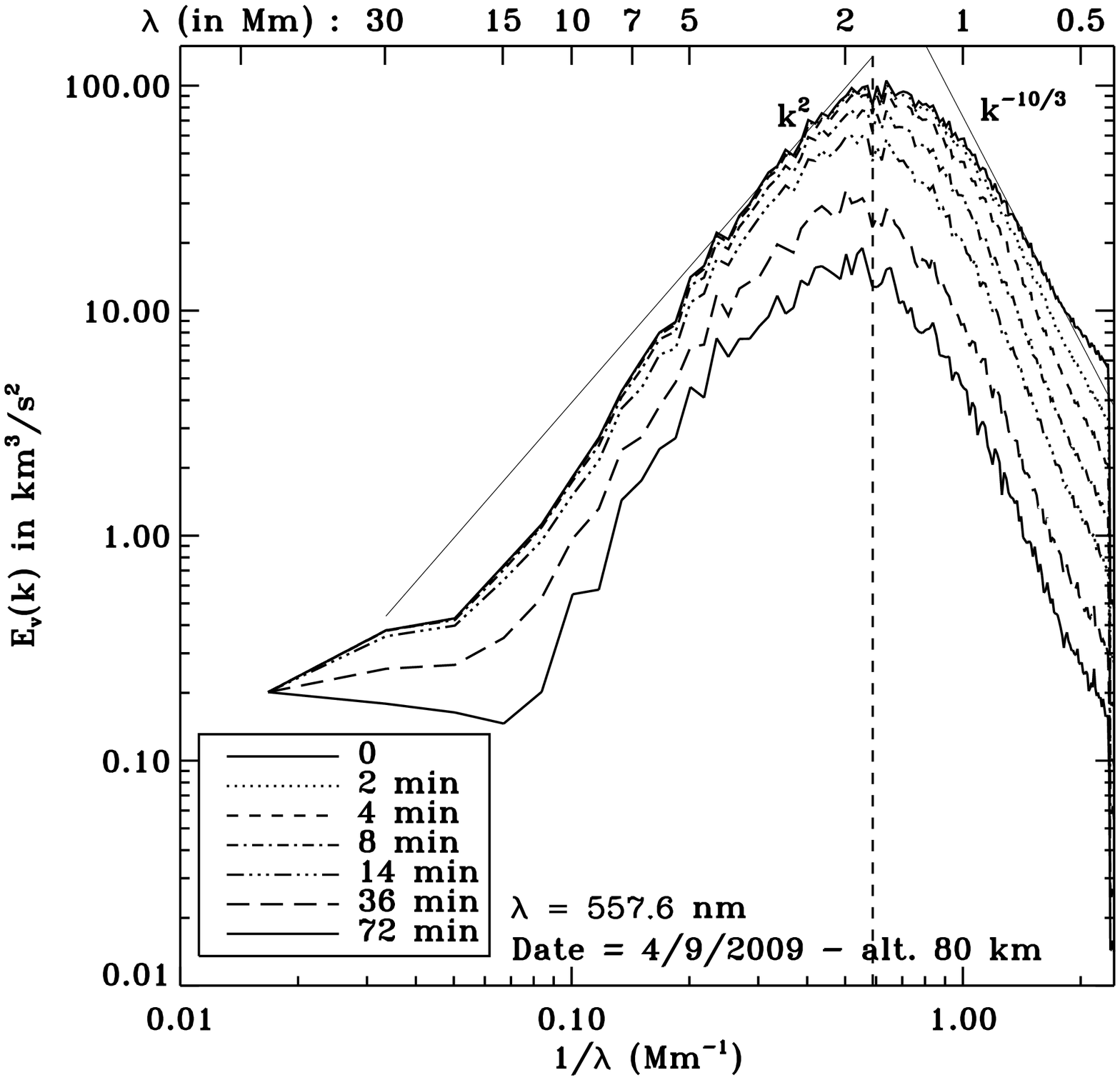}
\hspace*{10pt}
\includegraphics[width=0.48\linewidth]{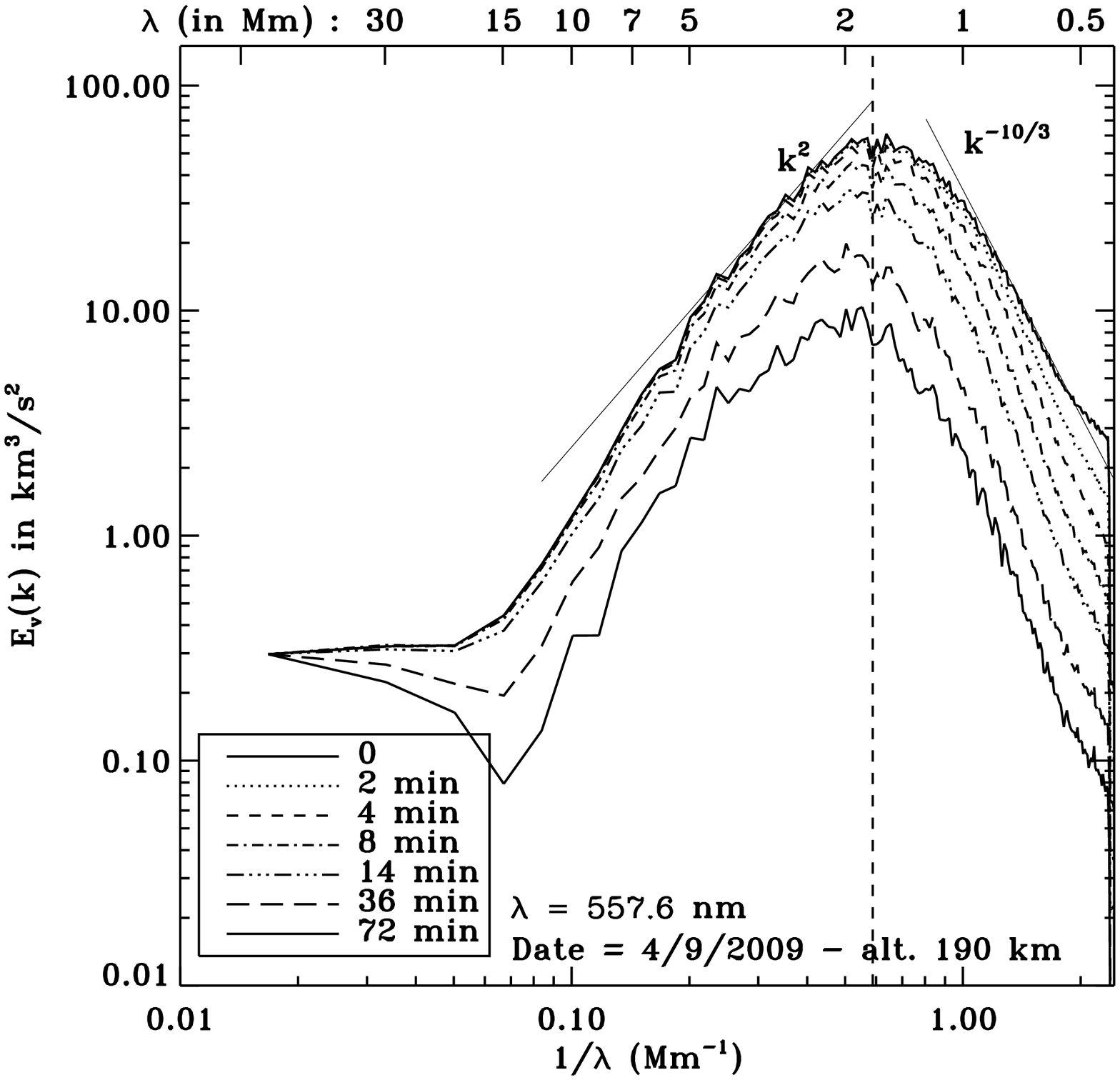}}
\caption[]{Kinetic energy spectra of the vertical velocity field for
various time windows. The vertical dotted line marks the maximum
of the spectrum at $\lambda=1.7$~Mm. Left: spectra computed with
Doppler shifts taken with a line chord of 16~pm, corresponding to an
mean altitude of 80~km. Right: same as left but for a chord of 8~pm,
corresponding to an mean altitude of 190~km. The original images were
flat-fielded according to the procedure described in Sect.~\ref{vv}.
Here $N_x=N_y=N_p=1024$.}
\label{kesv}
\end{figure*}

\subsection{Vertical velocity fields}

Let us now focus on the vertical velocity fields $v_z(x,y)$. As mentioned
above, these are obtained from the Doppler effect and thus each pixel of
the image gives a measure of this component of the velocity field.  We now
compute the horizontal spectral density of the vertical velocity, namely

\[ \demi\moy{\overline{v}_z^2} = \int_0^\infty E_v(k)dk\]
for various time averages.

\begin{figure}[htp]
\centerline{\includegraphics[width=0.9\linewidth]{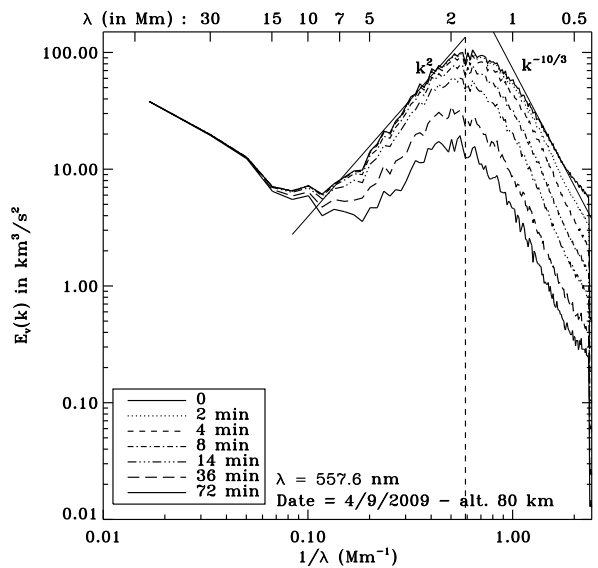}}
\caption[]{Same as in Fig.~\ref{kesv}- left, but with data not flat-fielded.}
\label{kesv_un}
\end{figure}

Figure~\ref{kesv} shows the spectra of the Doppler radial velocity
obtained with the bisector method after reconstruction of the line
profile at 557.6~nm. We extracted velocities at two different
altitudes. A line chord of 16~pm was used to sample a layer around an 
altitude of 80~km \citep{altrock_etal75,berrilli_etal02} and a second 
chord of 8~pm, deeper in the line, was used to sample the atmosphere 
around 190~km. In order to detect potential spurious effects associated 
with the artificial flat-fielding operation, an additional set of 
spectra was computed from images not flat-fielded (Fig.~\ref{kesv_un}).

First these spectra seem to follow $E_v(k) \sim k^2$
in the 2.5--7~Mm range. If the flat-fielding operation
is correct, this power law may extend up to 30~Mm. Note that this 
slope is reminiscent of the power law found from the relationship between
correlation time and the scale of horizontal velocities (see the
previous section and Fig.~\ref{KES}-right). In terms of velocity, the
$k^2$ power law for the spectral density implies $v_z\sim\sqrt{kE_v(k)}\sim
k^{3/2}\sim\lambda^{-3/2}$.

As noted above, the quality of the data allowed us to investigate
the subgranulation dynamics accurately for the first time. The absence of 
Earth atmospheric turbulence and the deconvolution of the images by the
instrumental transfer function leads to high quality horizontal spectra
of the vertical velocity. As shown by Fig.~\ref{kesv}, we observe in this
range of scales a decrease of the spectral density closely following a
$k^{-10/3}$ power law.  When the signal is slightly averaged over a
4-min interval, the cut-off is steeper and the exponent close to $-4$.

Concerning the granulation peak, the spectra show that the privileged
scale is 1.7~Mm. The spectral kinetic energy density is around
100~km$^3$/s$^2$ at a mean altitude of $\sim$80~km.  At an altitude around
190~km, the same spectra show an amplitude of 50~km$^3$/s$^2$.  This
reduction of the spectral kinetic energy density is likely associated with
the braking of upward motions which hit the stably stratified atmosphere.

\begin{figure*}[t]
\centerline{\includegraphics[width=0.48\linewidth]{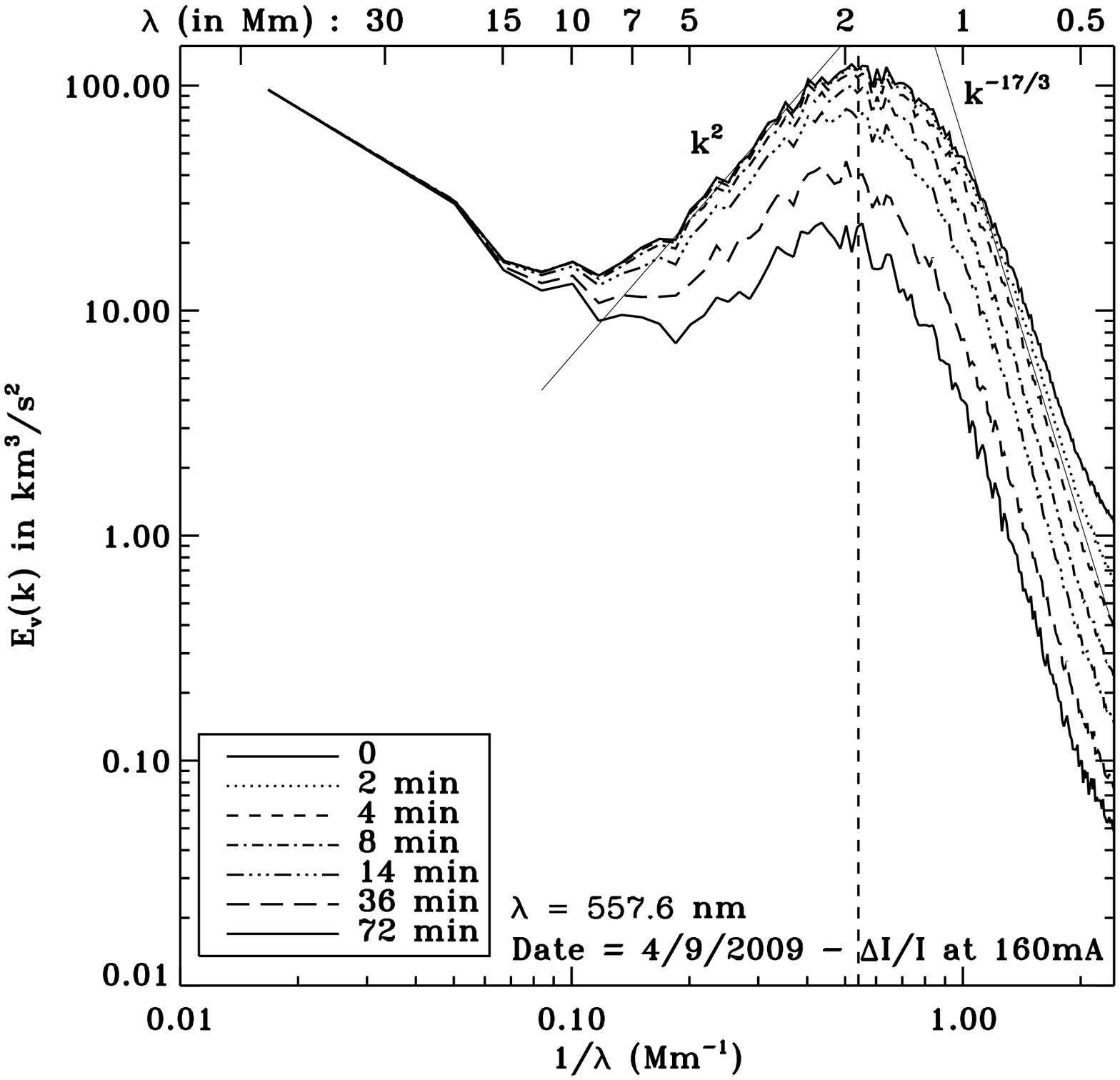}
\hspace*{10pt}
\includegraphics[width=0.48\linewidth]{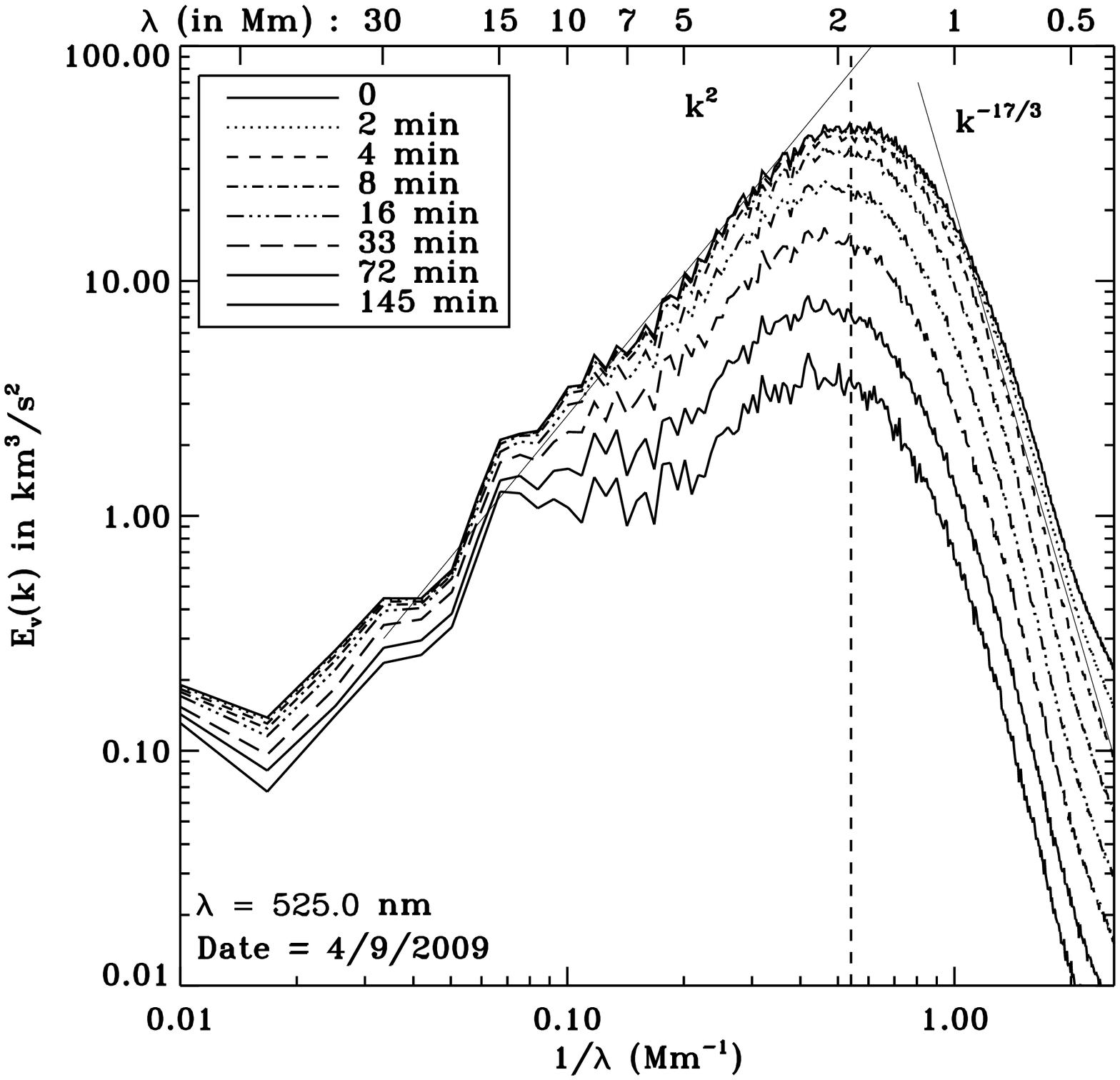}}
\caption[]{Left: same as in Fig.~\ref{kesv_un}, but with Doppler velocities 
computed with just two wavelengths in the line.
Right: same as left but for $\lambda=525.0$~nm (images are corrected using a
flat-field image provided by the Hinode mission, $N_x=N_y=704$ and $N_p=1024$).}
\label{kesv2}
\end{figure*}

To confirm these results with independent data, we used the third set
presented in Sect.~\ref{vv} taken with NFI around the FeI line at
525.0~nm. Unlike the previous data at 557.6~nm we only had access 
to two wavelengths around the FeI line and had to derive the Doppler
shift by the poor man's method, following Eq.~\ref{calib_eq}. The
results, shown in Fig.~\ref{kesv2}, are still interesting. 
With this method a $k^2$ power law is also obtained at large scales
both on the 557.6~nm and 525.0~nm data. The latter data, which 
benefit from a good flat-field image, also indicate that the $k^2$ power 
law extends up to 30~Mm.

However, we find that the cut-off is sharper on the subgranulation side,
not far from a $k^{-17/3}$ power law. This is presumably a residual of
the small-scale cut-off of the intensity power spectrum, which follows
the same power law (see below). This result suggests that the poor man's method 
is probably not appropriate for the computation of the small-scale 
radial velocity field.

\begin{figure*}[ht]
\centering
\includegraphics[width=0.48\linewidth]{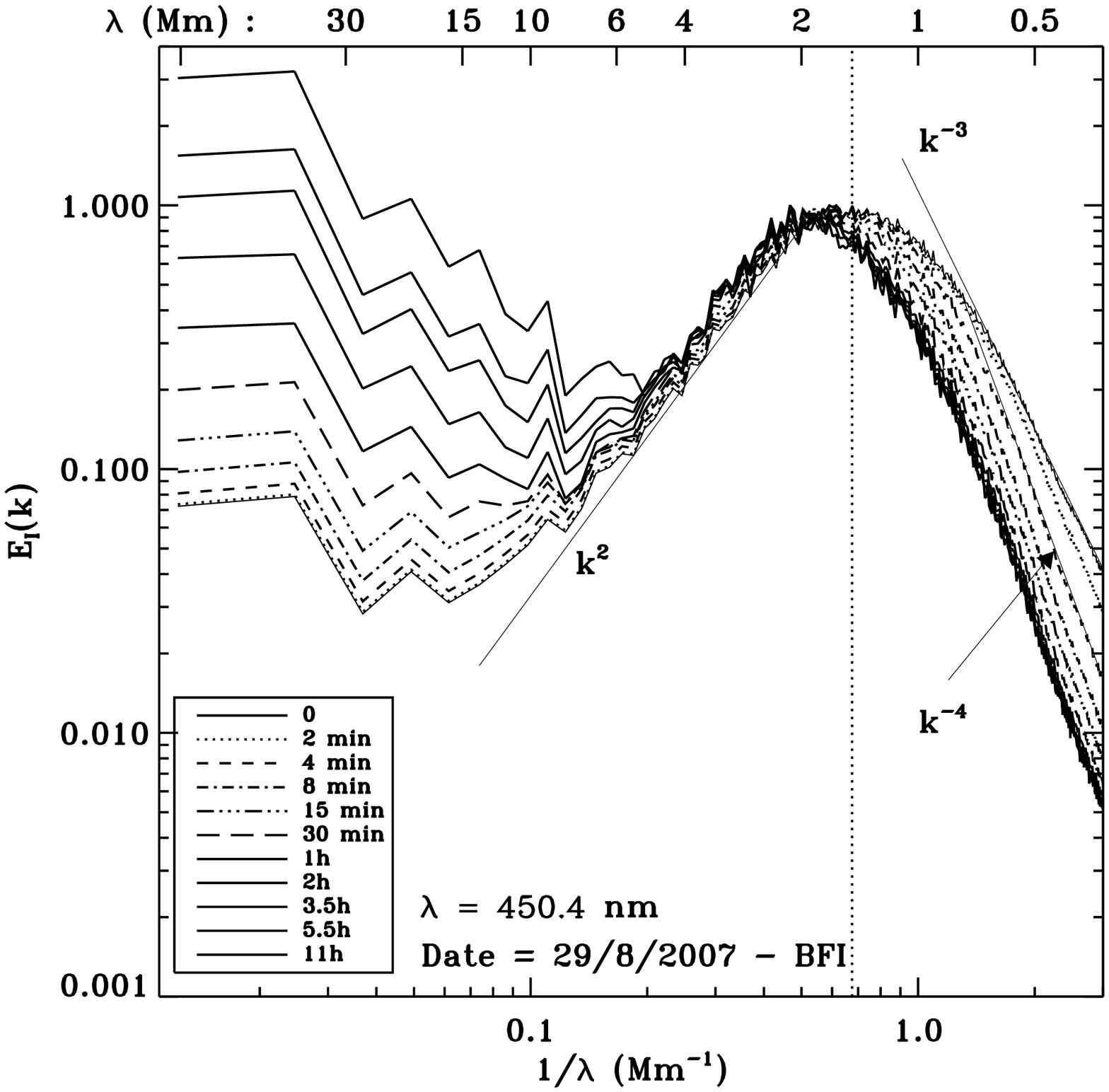}
\hspace*{10pt}
\includegraphics[width=0.48\linewidth]{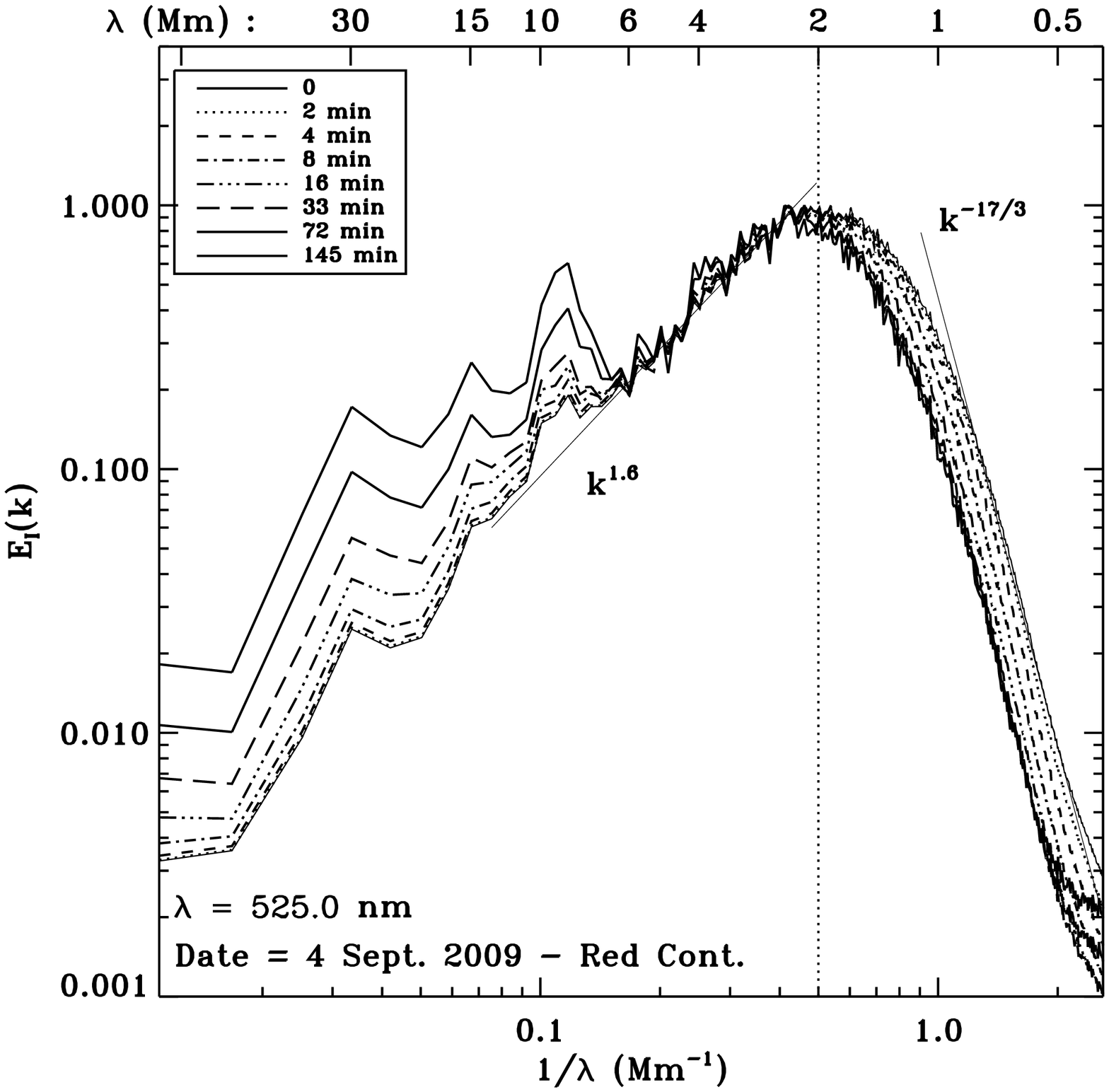}
\caption[]{Spectral energy density of the intensity field. Left: data
from the BFI at 450.4~nm (spectral width of 0.22~nm sampling the
blue continuum) for various time windows (data set of 29-31/8/2007). The
spectra are normalised so that the amplitude of the granulation peak is
unity. The vertical dotted line marks the position of the granulation
peak determined with instantaneous spectra. Here $N_x=941$, $N_y=934$
and $N_p=1024$. Right: data from the NFI at 525.0208+0.0108~nm (spectral
width of 9~pm sampling the continuum on the red side of the line); here
$N_x=N_y=704$ and $N_p=1024$.}

\label{spect_int}
\end{figure*}

\section{Intensity spectra\label{sectionfour}}

With the data sets presented in Sect.~\ref{vh} and \ref{vv} we
finally investigated the power spectrum of the horizontal intensity
fluctuations defined according to

\[ \demi\moy{\overline{\delta I}^2} = \int_0^\infty E_I(k)dk~.\]

The spectra obtained from the blue continuum images at $\lambda=450.4$~nm
are shown in Fig.~\ref{spect_int}. The variations in the mesoscale range
are characterised by a spectrum close to a $k^2$ power law, 
remarkably similar to the spectrum of vertical velocities.
Assuming that the surface radiates like a blackbody, these 
fluctuations reveal those of temperature. 

At scales smaller than a maximum located around the scale of 1.5~Mm, 
the power spectrum decreases steeply. In the subgranulation range, the
spectral density almost follows a power law in $k^{-3}$. We note that these
results agree with those of \cite{WR09} as derived from their Fig.~9. 
Our longer time series provides a better statistic, which probably improves 
the determination of the large-scale side of the spectra.

We then used the images taken at 525.0316~nm with the NFI,
whose width is 9~pm. It turned out that this light comes from the
continuum just besides the FeI line at 525.0208~nm \citep{uitenbroek09}. As shown 
in Fig.~\ref{spect_int} (right), the intensity spectra show a steeper cut-off
in the subgranulation range than for the previous data, close to a $k^{-17/3}$ law. 
We note that the granulation scale peak is at 2~Mm for this wavelength.

\begin{figure}[ht]
\centering
\includegraphics[width=0.90\linewidth]{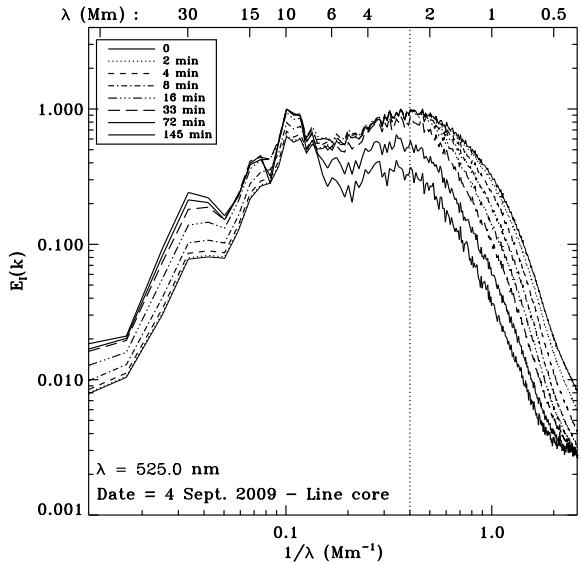}
\caption[]{Same as Fig.~\ref{spect_int} -right, but with the NFI almost 
centred on the line core at $\lambda=521.019$~nm.}
\label{spect_intL}
\end{figure}

Finally, the images taken at the blue side of the 525.0~nm FeI line are in
fact very close to the line core, namely at 525.019~nm. Atmospheric models
of \cite{uitenbroek09} suggest that this intensity emerges at an altitude
of 200~km above the continuum. The spectra shown in Fig.~\ref{spect_intL}
show a clear shift of the granulation peak towards larger scales,
namely 2.5~Mm. No power law arises in the meso-scales, while the
subgranulation range shows a roundish cut-off with no clear power law.

\section{Discussion and physical interpretation\label{discussion}}

\subsection{Large scales}

The foregoing results first confirm that the power spectra of horizontal
flows at the Sun's surface are dominated by two scales corresponding
to granulation at $\sim$1.7~Mm and supergranulation at $\sim$30~Mm. This result
confirms the one of \cite{HBBBKPHR00}, who used only one velocity component
(the radial velocity) of the flow field. As far as the vertical velocity is
concerned, the granulation scale clearly emerges at 1.7~Mm. On the large-scale
side of granulation, we found that the vertical velocity field decreases like 
$\lambda^{-3/2}$ with the horizontal length scale $\lambda$, corresponding to 
a $k^2$-law for the associated kinetic energy spectral density.  This law seems
to be valid up to the supergranulation scale since its prediction of a
supergranule vertical velocity corresponds to the value of 30~m~s$^{-1}$ 
found by \cite{HBHR02} with MDI/SOHO data. Note that this conclusion holds
only if the flat-field operations on the images did not damage the data.

The $k^2$-scaling law of vertical kinetic energy can be related to
the scaling law of the horizontal kinetic energy spectral density using
physical arguments. Indeed, mass conservation (see below Eq.~\ref{mc})
implies that $v_z/H\sim v_h/\lambda$, where $H$ is the vertical scale
height.  If $H$ is independent of $\lambda$, and since $v_z\sim
\lambda^{-3/2}$, then $v_h\sim\lambda^{-1/2}$, meaning that
$E_h(k)\sim$~const. This law is approximately consistent with the
horizontal flow data:  as shown by the 30-minutes spectrum in Fig.~\ref{KES},
the spectral density $E_h(k)$ does not vary very much in the 
2.5--20~Mm range, where it stays between 190 and 260 km$^{3}$/s$^2$.

To conclude on the $k^2$ power law, we note that it also appears
in the intensity fluctuation spectrum obtained with the  broadband filter
in the blue continuum. This looks like a strong correlation between
buoyancy fluctuations traced by intensity fluctuations and vertical
velocities. However, we observe that this scaling is not as conspicuous at
another wavelength with a narrower filter, which likely samples different
layers (see discussion below).

The derived horizontal flows may also be used to obtain some information
on the scale height of the variations of the vertical velocity at the
scale of supergranulation. Assuming that the anelastic approximation 
can be used at these scales (it means that acoustic waves are filtered out),
we can write the mass conservation as

\beq \partial_zv_z = -v_z\partial_z\ln\rho -\vec{\na}_h\cdot\vv_h\, ,\eeqn{mc}
where $\vv_h$ is the horizontal velocity field and $\rho$ is the
background density.  Using the simple model of \cite{STW91} for
a horizontal divergent source, namely $v_r=vr/R\exp(-r^2/R^2)$, we
could estimate for a well-formed supergranule that the divergence is
approximately $10^{-4}$~s$^{-1}$. From \cite{SN98}, $\partial_z\ln\rho\sim
2.4$~Mm$^{-1}$ near the surface. Then, using the estimate of 30~m~s$^{-1}$ 
for the vertical velocity at this scale, we find that the vertical scale
height is $\sim1$~Mm, which indicates that the supergranulation flow is
likely shallow. This qualitative estimate agrees with recent 
local helioseismic inferences based on Hinode data \citep{Sekii_etal07},
which indicate that the imprint of supergranulation flows disappears 
around 3~Mm below the surface. Note that the foregoing exercise has
been done using typical values of all quantities and just gives a first
impression. To go further and get a more reliable view, this calculation
should be applied to a large sample of supergranules.

\subsection{Sub-granulation scales: an injection-diffusive range ?}

\begin{figure}[t]
\centering
\includegraphics[width=0.99\linewidth]{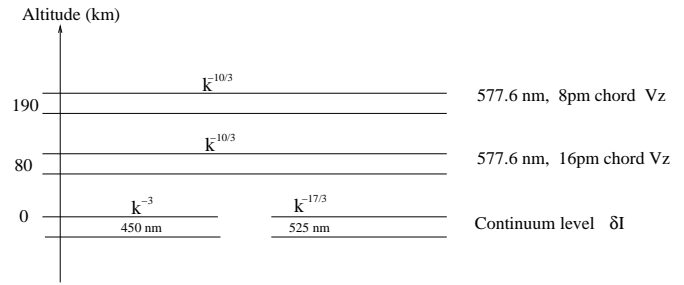}
\caption[]{Schematic view of the results. We indicate the layers where power
laws are perceptible in the subgranulation range.}
\label{sumview}
\end{figure}

On the other side of the granulation peak, in the subgranulation range,
the power spectrum of vertical velocity closely follows a $k^{-10/3}$ law, 
which steepens to $k^{-4}$ with time averaging, while the intensity variance
spectrum seems to follow either a $k^{-3}$ law or a $k^{-17/3}$
law depending on the data set used. These results are summarised in 
Fig.~\ref{sumview}.

A natural question to ask is what these scaling laws tell us about
the physics of the solar surface. It is indeed well established that a
variety of characteristic power law regimes can be derived for all kinds
of turbulent complex flows depending on the detailed physical processes at
work in the flow (e.g. shear turbulence \citep{lumley67,lohse96}, stably
and unstably stratified turbulence \citep{Obou59,Bol59,Bol62,lvov91}, MHD
turbulence \citep[][ and references therein]{schekochihin08}) and on the
relevant range of scales (e.g. injection, inertial or dissipation range)
for each physical field in the problem.  In the paragraphs below,
we therefore attempt to determine if the observed subgranulation scaling
exponents betray a particular flow regime.

First of all, considering the many differences between homogeneous,
isotropic, incompressible turbulence and solar surface turbulence at
observable scales \citep{NSLT97,petrovay01,rincon07}, there is no real
reason to expect that the Kolmogorov phenomenology of turbulence (which
predicts a $k^{-5/3}$ power law for the velocity field in the inertial
range) applies in the vicinity of the granulation scale.

A possible interpretation for the joint observation of the $-10/3$
and $-17/3$ exponents is in terms of buoyant dynamics in the injection
range of solar surface turbulence (i.e. not in the inertial range).
Let us assume that (i) buoyancy fluctuations are directly proportional
to temperature fluctuations (i.e. pressure fluctuations are not
influential) and that (ii) the dynamic is dominated by a balance between
inertial terms (or acceleration) and buoyancy forcing in the momentum
equation. Further assuming that velocity and temperature perturbations
have the following horizontal scale $\lambda$ dependence

\[ v_h(\lambda) \sim \lambda^\alpha, \qquad v_z(\lambda) \sim \lambda^\beta,
\qquad \delta T(\lambda) \sim \lambda^\gamma\]
and that vertical variations are controlled by the typical pressure or
density scale height $H$ independent of $\lambda$,  mass conservation
imposes that $v_z/H\sim v_h/\lambda$, implying $\alpha=\beta+1$.
Assuming finally the locality of non-linear interactions, (i) and (ii)
translate into the dynamical balance

\[ v_h v_z/\lambda + v_z^2/H \sim \delta T~,\]
hence $\gamma=2\beta$.  Now, if the horizontal fluctuations
of temperature follow a $k^{-17/3}$ spectrum, i.e. $\delta T\sim
\sqrt{\lambda^{17/3-1}}\sim \lambda^{7/3}$, then $\gamma=7/3$, $\beta=7/6$
and the horizontal spectrum of the vertical velocity is $E_v(k)\sim
k^{-10/3}$, just as observed\footnote{Assuming isotropy, the dynamical
balance would instead read $v_z^2/\lambda\sim \delta T$, leading to
$E_v(k)\sim k^{-13/3}$ for $E_I(k)\sim k^{-17/3}$.This argument can be
formalised using a generalised exact von K\'arm\'an-Howarth-Kolmogorov
equation for thermal convection \citep{rincon06}.}. Hence it is possible
to relate the horizontal fluctuations of intensity at the continuum level
to those of the vertical velocity in the upper layers (80~km or 190~km above)
in the subgranulation range.

Still, this is only one half of the story, since the argument does not 
predict the $-17/3$ slope for the intensity spectrum (it uses it as an input). 
This is because two dominant balances are required to determine scaling laws 
consistently in active scalar turbulence.  In the context of thermal convection, 
the second balance is obtained by isolating the dominant physical effects
in the temperature equation or, to be more precise, in the Yaglom equation
\citep{rincon06}, which is the analog of the von K\'arm\'an-Howarth
equation for the statistics of scalars.  In the Boussinesq approximation
(which is not suitable for the problem of solar convection), a $k^{-17/3}$
spectrum for temperature fluctuations can only be obtained if (i) the
following low P\'eclet number dominant balance\footnote{The P\'eclet number
of a thermal eddy is the ratio between its thermal diffusion time and
its typical turnover time. It is effectively of the order of one or smaller in
the sub-granulation range.}

\begin{equation}
\label{eq:temp}
v_z\,\partial_z\moy{T}\sim\kappa\,\Delta \delta T
\end{equation} 
is assumed in the temperature equation (where $\kappa$ is the
thermal diffusivity), i.e. temperature fluctuations are the diffusive
response to velocity fluctuations advecting a mean, slowly varying
temperature profile $\moy{T}$ and (ii) the velocity fluctuations
are not influenced by temperature fluctuations and have a $k^{-5/3}$
spectrum (Kolmogorov inertial range), i.e. $v_z \sim \lambda^{1/3}$.
Then, $\delta T\sim \lambda^2v_z \sim \lambda^{7/3}$, or equivalently
$E_I(k)\sim k^{-17/3}$.  This is the so-called inertial-diffusive regime
\citep[see e.g.][]{Lesieur90}. Note that for any velocity spectrum slope
steeper than -5/3 (say -10/3) Eq.~(\ref{eq:temp}) predicts a spectrum
much steeper than $k^{-17/3}$ for temperature fluctuations.  Hence,
it remains to be understood at this stage how the two slopes observed at
the solar surface can be made fully consistent using dynamical arguments.

Another interesting result regarding the sub-granulation range dynamics
is that the intensity fluctuations taken with the BFI in the ``blue
continuum" at 450~nm have a much weaker spectral cut-off close to a
$k^{-3}$ power law ($k^{-4}$ if slightly averaged) than those taken in
the continuum near the FeI line at 525~nm.  This is quite surprising,
because the fluid layer contributing most to the blue continuum should
be quite similar to the one contributing to the 525~nm continuum from
which the $k^{-17/3}$ spectrum was derived.  Beside the simple wavelength
difference, which indicates a slighly lower contributing layer for the
450~nm radiation, the most obvious difference between the two signals
comes from the spectral width of the filters. The width of the NFI
of 9~pm is such that the continuum near 525~nm is ``pure", while the
width of the BFI at 450.45~nm is 220~pm and includes a few absorption
lines. For this reason, the layers contributing to the light collected
in the two wavebands could be somewhat different. At the moment we have
no obvious explanation of these two different cut-offs for the intensity
spectrum in the continuum.  Of course we cannot totally exclude some
unexpected bias in the data processing, but if the result is confirmed
by independent approaches, detailed numerical simulations of the dynamics
are likely the only way to elucidate this puzzle.

For the sake of completeness, we mention that a $k^{-3}$ spectrum could
be a simple signature of a physically smooth temperature field, which can
be expanded in Taylor series. Indeed, if $T(\lambda)=T(0)+\lambda\na_h
T_{\left|{\lambda=0}\right.}$ then $\delta T = T(\lambda)-T(0)\sim
\lambda$, or equivalently $E_I(k)\sim k^{-3}$.  But, similarly to the
argument presented previously, we underline that this scaling is not fully
consistent with a $k^{-10/3}$ scaling for the velocity field if a dominant
balance between inertia and buoyancy is assumed in the momentum equation.

The previous theoretical arguments should of course not be taken at
face value.  They simply suggest that it might be possible to construct
some sort of ``injection-diffusive" turbulent regime characteristic of
the sub-granulation range, and serve to illustrate that the standard
framework of incompressible homogeneous isotropic turbulence is probably
too restrictive to theoretically explain the observations of the thermal
boundary layer at the solar surface.

These new observational measurements for the subgranulation range
definitely call for a deeper investigation.  A detailed theoretical
understanding of the solar power spectrum at the beginning of the
subgranulation range 0.1--1~Mm might be obtained by re-analysing existing
numerical simulations of solar granulation \citep{SN98}, looking for
the dominant terms in the temperature (or energy) equation\footnote{
It is remarkable that some of the early numerical
simulations by \cite{NSLT97} seemed to produce a $k^{-17/3}$ scaling
for intensity fluctuations.}.  Another possibility would be to perform
idealised high-resolution numerical simulations of thermal turbulence
done in the low-P\'eclet approximation \citep{thual92,lignieres99}
to better understand this complex and fairly unusual turbulence
regime from the theoretical point of view.

\section{Conclusions\label{conc}}

The power spectrum of surface flows at the Sun's surface has
been investigated using mainly image series from the Hinode/SOT
instrument. The results may be summarised as follows.

\begin{enumerate}
\item The smallest scales, i.e. the subgranulation range, can be
investigated for the first time without a seeing effect and using
corrections of the MTF of the instrument. On the one hand, the
spectral cut-off of intensity fluctuations is similar to a $k^{-17/3}$
power law for images taken with a narrow filter (width $\sim9$~pm) in
the continuum on the red side of the 525~nm FeI line and to $k^{-3}$
for images taken with the broadband filter (width $\sim220$~pm) around
450.45~nm.  On the other hand, the vertical velocity horizontal spectrum
measured 80~km or 190~km above the continuum at 557.6~nm shows a steep
cut-off in $k^{-10/3}$, which can be related to the  $k^{-17/3}$ power
law by invoking a balance between buoyancy and inertia. The values of
the two scaling exponents suggest that the temperature fluctuations
impact the dynamics at a scale well below the one determined by the unit
P\'eclet number.  Finally, we also note the decrease of the granulation
peak amplitude with altitude, possibly betraying the deceleration of
the upflows.

\item The granulation spectral peak at 1.7~Mm is conspicuous in the
vertical velocity spectrum. The location of the intensity spectrum peak
depends on the wavelength: it is close to 1.5~Mm in the blue continuum at
450~nm, 2~Mm in the continuum near 525~nm, and to 2.5~Mm in the core of
the FeI line at 525.0~nm, which probes the dynamics in a layer 200~km above
the continuum.

\item Moving to the large scales, we note that the spectral density
of vertical kinetic energy decreases as $k^2$ with decreasing $k$
(equivalently, the amplitude of $v_z$ decreases as $\lambda^{-3/2}$
with the horizontal scale $\lambda$). This $k^2$ power law also emerges
from the relationship between the scale and the correlation time of the
horizontal flows.  Furthermore we note that
the spectral density of intensity fluctuations in the blue continuum
also scales like $k^2$ in the 2.5--10~Mm range.

\item The power spectrum of horizontal flows has been explored in
the 2.5--40~Mm range.  It clearly reveals the supergranulation peak
as well as the large-scale side of the granulation peak.
These observations confirm in passing the result of the numerical simulations of
\cite{RRLNS01} that granule tracking cannot determine plasma flows at
scales below $\sim$2.5~Mm. The energy density minimum is found
at a scale of 12~Mm. These results confirm with a completely different
technique the spherical harmonics spectra obtained by \cite{HBBBKPHR00}
from dopplergrams. Focusing on the supergranulation peak, we could
estimate its amplitude to be in the range of 200--700~km$^3/$s$^2$ and show
its strong variability. From Figs.~\ref{KES} and \ref{pore} we may add
that the visible part of the granulation peak points to an amplitude in the
range of 200--400~km$^3/$s$^2$. Interestingly enough, we discovered that
the emergence of a pair of pores is able to erase the supergranulation
peak completely, which points out the sensitivity of the supergranulation
amplitude flow to the ambient magnetic field. Finally, combining all the
data, we tentatively estimated the scale-height of the vertical velocity
field at the supergranulation horizontal scale. We found a scale height
of approximately 1~Mm, indicating that supergranules are likely surface
flows. However, this estimate should be verified for a number
of supergranules to obtain a statistically robust value.

\end{enumerate}

These observational results provide a set of landmarks that should be
recovered by future numerical simulations. Future observational work will 
focus on the large-scale side of the supergranulation peak to 
determine what scaling law, if any, controls the very large-scale dynamics 
of the Sun between supergranulation and differential rotation. Another point 
deserving a detailed investigation is the dynamical influence of the 
magnetic fields on the surface flows and their power spectrum. 

\begin{acknowledgements}
We are grateful to the Hinode team for the possibility to use their data.
Hinode is a Japanese mission developed and launched by ISAS/JAXA,
collaborating with NAOJ as a domestic partner, NASA and STFC (UK)
as international partners. Scientific operation of the Hinode mission
is conducted by the Hinode science team organised at ISAS/JAXA. This
team mainly consists of scientists from institutes in the partner
countries. Support for the post-launch operation is provided by JAXA
and NAOJ (Japan), STFC (U.K.), NASA, ESA, and NSC (Norway).This work
was (partly) carried out at the NAOJ Hinode Science Center, which
is supported by the Grant-in-Aid for Creative Scientific Research
``The Basic Study of Space Weather Prediction" from MEXT, Japan (Head
Investigator: K. Shibata), generous donations from Sun Microsystems,
and NAOJ internal funding.

We thank Fran\c{c}ois Ligni\`eres for several interesting comments and
remarks on the manuscript. We are also very grateful to Han Uitenbroek
who provided us with simulations results of line formation in the
Sun's atmosphere.  This work was supported by the Centre National de
la Recherche Scientifique (C.N.R.S., UMR 5572), through the Programme
National Soleil Terre (P.N.S.T.).

\end{acknowledgements}

\bibliographystyle{aa}
\bibliography{../../biblio/bibnew}

\appendix
\section{Absolute spectra\label{app}}

In this paper, we give the kinetic energy density spectra with their
absolute values in km$^3$/s$^2$. To facilitate future comparisons, we
document the way these spectral densities are computed in detail. We
recall that the kinetic energy spectrum $E(k)$ is defined as

\begin{equation} \demi\moy{v^2} = \int_0^\infty E(k)dk\, .\end{equation}
It is related to the Fourier transform of the velocity components
by

\begin{equation} \moy{\hat{v}_i^*(\vk)\hat{v}_j(\vk')} = \phi_{ij}(\vk)\delta(\vk-\vk')\,
, \end{equation}
where $\phi_{ij}(\vk)$ is the Fourier transform of the
two-point correlation function of the velocities, namely, $Q_{ij}
(\vr)=\moy{v_i(\vx)v_j(\vx+\vr)}$. $\delta$ is the Dirac distribution.
In two dimensions (our case here), $E(k)$ and $\phi_{ij}$ are related by

\begin{equation} E(k) = \demi k\int_{0}^{2\pi} \phi_{ii}(\vk)d\theta_k\, ,\end{equation}
where $\theta_k$ is the polar angle of $\vk$. The numerical 2D-Fourier
transform (like in the IDL software) is defined as

\begin{equation} \tilde{f}=\frac{1}{N_xN_y}\sum_{x,y}f(x,y)e^{-2i\pi(k_xx + k_yy)}\, ,\end{equation}
which means that the Dirac distribution in a discrete way reads

\begin{equation} \tilde{\delta}=\frac{1}{N_xN_y}\sum_{x,y}e^{-2i\pi
(k_xx + k_yy)}\, ,\end{equation}
Now, spectral quantities in turbulence theory usually use
\begin{equation} \hat{f}(\vk) = (2\pi)^{-2}\int f(\vr)e^{-i\vk\cdot\vr}dxdy\end{equation}
as the definition of the 2D-Fourier transform. Setting $\vk=2\pi\vk'$ then

\begin{equation} \hat{f}(\vk) = (2\pi)^{-2}\int f(\vr)e^{-2i\pi\vk'\cdot\vr}dxdy\, ,\end{equation}
The correspondence is accordingly

\begin{equation} \hat{f}(\vk) = \tilde{f}(\vk')N_xN_y p^2/(2\pi)^2 \quad
{\rm and}\quad \delta(\vk) = \tilde{\delta}N_xN_y p^2/(2\pi)^2\, ,\end{equation}
where $N_x$ and $N_y$ are the size of the data. $p$ is the linear size
of the pixel. We wish to derive the correct normalisation of $E(k)$
which is that of $\phi_{ii}(\vk)$. We note that $\phi_{ii}(\vk)$ is
the ratio of the squared norm of the Fourier components of the velocity
divided by the Dirac distribution. Using the preceeding normalisation for
both $\hat{v}$ and $\delta(\vk)$, we find that

\begin{equation} \phi_{ii}(\vk) = \frac{|\hat{v}|^2}{\delta(\vk)}
=\frac{|\tilde{v}|^2(N_xN_y)^2p^4(2\pi)^2}{(N_xN_y)p^2(2\pi)^4}=
|\tilde{v}|^2N_xN_yp^2/(2\pi)^2\; .\end{equation}
If we restrict our case to a square data set ($N_x=N_y=N$), then

\[ E(k)=\demi\int
\phi_{ii}(\vk)kd\theta_k=\demi\sum_{\forall
k'\in[k,k+dk]/2\pi}|\tilde{v}(k')|^2
\frac{2\pi}{N p}\frac{N^2p^2}{(2\pi)^2}\, ,\]
where we noticed that the dimensional wavenumber $k$ and the
non-dimensional one are related by a factor $2\pi/Np$. Finally,

\begin{equation} E(k) = \demi\frac{N p}{(2\pi)}\sum_{\forall
k'\in[k,k+dk]/2\pi}|\tilde{v}_x(k')|^2+|\tilde{v}_y(k')|^2\, ,\end{equation}
where $\tilde{v}_x$ and $\tilde{v}_y$ are the discrete Fourier
transform of $v_x$ and $v_y$.  Note that the summation $\sum_{\forall
k'\in[k,k+dk]/2\pi}$ implies that the $k'$ factor of the surface element
$k'd\theta_kdk'$ is automatically taken into account.  Sometimes the
quantity $P(k)=\sqrt{kE(k)}$ is used; here

\begin{equation} P(k)=\sqrt{\frac{1}{2}\sum_{\forall
k'\in[k,k+dk]/2\pi}k'|\tilde{v}|^2}\, .\end{equation}

\bigskip
\noindent{\bf Effect of zero padding:}
the Fourier transforms are done on a square of size $N\times N$, while
the data usually occupy a rectangle of the size of $N_x\times N_y$ ($N_x\leq
N, N_y\leq N$); the energy of the Fourier modes is thus a fraction
$N_xN_y/N^2$ of the total. Hence, to obtain spectra which are independent
of the zero padding, we need to multiply the resulting spectral densities
by $N^2/(N_xN_y)$ so that

\begin{equation} E(k) = \frac{N^3 p}{4\pi N_xN_y}\sum_{\forall
k'\in[k,k+dk]/2\pi}|\tilde{v}|^2\, ,\end{equation}

\bigskip
\noindent{\bf Effect of the normalisation of the Fourier
Transform:} if the Fourier Transform is multiplied by a factor $A$, then $E(k)$ is
multiplied by $A$ and $P(k)$ is multiplied by a factor $\sqrt{A}$. The 
$P(k)$ spectrum has no absolute significance.

\end{document}